\documentclass[english,prl,preprint]{revtex4}
\usepackage[T1]{fontenc} \usepackage[latin9]{inputenc}
\usepackage{verbatim} \usepackage{booktabs} \usepackage{amsmath}

\makeatletter

\providecommand{\tabularnewline}{\\}

\usepackage{amsfonts} \usepackage{dcolumn} \usepackage{bm}
\usepackage{graphicx}

\newcommand{\degree}{\ensuremath{^\circ}}

\usepackage{babel} \makeatother

\begin{document}

\title{Mirror symmetry rupture in double photoionization of endohedrally confined atoms}
\author{F. D. Colavecchia$^{1}$, G. Gasaneo$^{2}$, and
  D. Mitnik$^{3}$} \affiliation{$^{1}$Div. Colisiones Atómicas,
  Centro Atómico Bariloche and Conicet, 8400 S. C. de Bariloche, Río
  Negro, Argentina} \affiliation{$^{2}$Departamento de F\'{\i}sica,
  Universidad Nacional del Sur and Conicet, 8000 Bah\'{\i}a Blanca,
  Argentina} \affiliation{$^{3}$Instituto de Astronomía y Física del
  Espacio, Dep. de F\'{\i}sica, Universidad de Buenos Aires, and Conicet, 1428 Buenos Aires, Argentina}

\date{\today}

\begin{abstract}
  We study the double electronic emission by photon impact from He in the
  center of a spherical fullerene, which is modeled by a square-well shell.
  This system exhibits a manifold of avoided crossings as a function of the 
  well depth, and 
  present mirror colapses. However, this symmetry
  is broken in the triple differential cross section due to
  the delocalization of the He electrons in the initial state.
  Moreover, the fullerene potential involves higher angular momenta partial waves to
  be included in the process, which modifies the well-known
  two-lobe cross section from isolated He.

\end{abstract}
\maketitle
Fullerene molecules, formed by pentagonal and hexagonal arrangements
of atoms, have received attention right from its discovery and
undoubtely opened many new captivating areas in physics. These
molecules come in different sizes and shapes, from the well known,
quasi spherical $\textrm{C}_{60}$ carbon
fullerene \cite{kroto_c60:_1985} to nanotubes.  Beyond the specific
features of fullerenes, many methods to include atoms inside their
shells have been
developed \cite{bethune_atoms_1993,yannoni_scandium_1992}.  These
compounds form stable molecules that also attracted the interest of
the community, such as their potential use as nanocages to storage
atoms  \cite{pupysheva_fullerene_2008,footn1} . Many
properties of these systems are determined by the differences in the
physics of the embedded atom compared to the isolated one.

Some of the most interesting of these features arise when these
endohedrally embedded atoms interact with light and electronic
emission occurs during the process. Since the pioneering work of Puska
and Nieminen \cite{puska_photoabsorption_1993}, different kinds of
resonant behavior have been identified. First evidence of these
effects were observed in the photoionization of $\textrm{C}_{60}$,
revealed as oscillations in the cross
sections \cite{xu_oscillations_1996} and attributed to the ability of
the carbon shell to support an intramolecular standing
wave. Confinement resonances are predicted for endohedrally embedded
atoms such as
$\textrm{Xe@C}_{60}$ \cite{amusia_strong_2000,amusia_confinement_2004},
due to the reflection of the photoelectron in the fullerene cage.
Furthermore, the role of multielectronic correlation has been
investigated recently, giving rise to the so-called correlation
confinement resonances \cite{dolmatov_correlation_2008}, and also the
interference among resonances \cite{amusia_photoionization_2008-1}.
However, all these works involved a complex multielectronic atom
inside the fullerene shell, that gives rise to many mechanisms of
electronic emissions even when the atom is isolated.

Few works are devoted to multiple photoionization of
fullerenes or endohedrally embedded atoms. Kidun et al. have examined
the multiple photoionization of $\textrm{C}_{60}$ within a many
particle approach \cite{kidun_multiple_2006}.  Two-electron
photoionization has been studied for high photon energies by
Amusia \cite{amusia_two-electron_2006}, but only the double-to-single
photoionization cross sections ratio were reported. They analized the
process with highly asymmetrical energy sharing, where the energy
of one electron is much bigger that the energy of the
other one. The emission in that case proceeds with a shake off
mechanism, where the fast electron absorbs most of the energy of the
photon, while the slow one is ejected due to the residual modified
atomic field left after the ejection of the first one. Since their
model of the fullerene potential was a delta function, it did 
not take into account accurately the possible
delocalization of the atomic electrons into the cage.

Whereas the investigation of bound or continuum states of two electron
systems can be performed by different techniques, double
photoionization (DPI) of atomic or molecular species by single photons
has unique advantages. First, it enables to directly probe the
dynamics of the electron pair both in the initial and final states with the
same collisional process. Besides, it is free of long-range correlations
between the target and the incoming particles. Finally, the complete
absorption of the photon by both electrons is determined by the
interelectronic correlation. Then, it is not surprising that this
process has been thoroughly studied along the years for both atomic
and molecular species \cite{avaldi_photodouble_2005}.

In this work we bring a different perspective to the problem of DPI
from endohedrically embedded atoms. We choose the simplest system
to study the role of the fullerene cage in the electronic emission,
as well as the influence of the interelectronic correlation in
the process, both in the bound state as well as in the continuum
one. To this end, we consider a He atom in the
center of the fullerene cage and assume that there are only two
active electrons in the system. We also focus on equal energy sharing conditions for the
ejected photoelectrons. The simplified electronic structure of the
fullerene molecule seen by an electron has been usually described
through a cage model potential   \cite{puska_photoabsorption_1993}: \begin{equation}
  V_{w}(r)=\left\{ \begin{aligned}-U_{0}\quad & \textrm{if}\quad r_{c}\leq r\leq r_{c}+\Delta\\
      0\quad & \textrm{otherwise}\quad\end{aligned}
  \right.\label{vcage}\end{equation} For C$_{60}$, $r_{c}=5.75$ a.u.,
$\Delta=1.89$
a.u.   \cite{puska_photoabsorption_1993,xu_oscillations_1996}.
These simple assumptions enable one to discover all the richness of
these systems. Moreover, different spherical fullerenes are described 
varying the well depth. In fact, the energetic structure
of these molecules as a function of the magnitude of the cage potential
$U_{0}$ presents a manifold of avoided crossing between states, even
within this two-electron model  \cite{mitnik_endohedrally_2008,neek-amal_ground-state_2007}. 
Similar results were found for
endohedrally embedded hydrogen  \cite{connerade_electron_1999}.  For
the sake of simplicity, we choose to analyze the crossing between the
ground and first excited states. Thus, the initial state of the system
will be written in terms of one electron 1s ($\phi_{1s}(r)
=4\sqrt{2}\exp{(-2r)}$) and 2s ($\phi_{2s}(r) =2\exp{(-r)}(1-r)$)
states of the isolated atom, and the approximate solution of the model
potential (\ref{vcage}), $
  \phi_{w}(r)=\exp{(-\alpha(r-r_{0})^{2})} $
where $r_{0}=r_{c}+\Delta/2$. We construct the following configuration
interaction (CI) wave function with these one-electron wave functions
for the bound state: \begin{equation}
  \begin{split}
    \Psi_{i}(\mathbf{r}_{1},\mathbf{r}_{2})&=N  \left\{ a\;\varphi_{1s}(r_{1})\varphi_{1s}(r_{2})\varphi_{corr}(r_{12})\right.\\
    & +\left.b\;\left[\varphi_{1s}(r_{1})\phi_{w}(r_{2})+1\leftrightarrow2\right]\right.\\
    &
    +\left.c\;\left[\varphi_{1s}(r_{1})\varphi_{2s}(r_{2})+1\leftrightarrow2\right]\varphi_{corr}(r_{12})\right\}
  \end{split}
  \label{eq:ci}\end{equation}
where $\varphi_{corr}(r_{12})=1+r_{12}/2$ is a correlation factor.

%

This state is suitable to analyze the collision process near the
crossing between the ground and the first excited state of the system
as a function of the potential depth. The basis parameters ${a,b,c}$
as well as the exponential factor $\alpha$, the normalization constant
$N$ and the energy depend on $U_{0}$ and are obtained with 
variational techniques, see Table 1. The crossing is at
$U_{0}^{cross}=1.35$ a.u.\cite{footn2} We also choose two other values of the magnitude of the cage
potential $U_{0}$ to calculate the cross sections, one below
($U_{0}=1.1$ a.u.) and one above ($U_{0}=1.65$ a.u.) the crossing of
the levels. This particular selection of the basis does enhance the
interplay between the free atomic ground state and the partially
delocalized state with one electron in the atom and the other one in
the fullerene, since the contribution of the $1s2s$ state through
coefficient $c$ is much smaller than the other ones. Thus, the basis
parameters measure the degree of localization of the electrons: when
$a\approx1,$ both electrons are near the atomic core, while
$b\approx1$ implies that the electronic density spreads up to the
fullerene cage.

We assume a dipolar approximation for the DPI because usually
experimental data is obtained in collisions with low energy
photons  \citep{avaldi_photodouble_2005}.  The triply differential cross
section (TDCS) in terms of the momenta $\mathbf{k}_{1}$ and
$\mathbf{k}_{2}$ of the two ejected electrons can be obtained in the
velocity gauge in terms of the transition matrix $
T^{(V)}(\mathbf{k}_{1},\mathbf{k}_{2})=\left\langle
  \Psi_{f}^{-}\left(\mathbf{r}_{1},\mathbf{r}_{2}\right)\left\vert
    \mathbf{\varepsilon}\cdot\left(\nabla_{1}+\nabla_{2}\right)\right\vert
  \Psi_{i}\left(\mathbf{r}_{1},\mathbf{r}_{2}\right)\right\rangle $
where $\mathbf{\varepsilon}$ is polarization of the incoming light.
We recall that this calculation can be performed also in acceleration
or length gauges. Each gauge form emphasizes different regions of the
configuration space, but all of them should give the same theoretical
description of the process, provided that both the initial
$\Psi_{i}\left(\mathbf{r}_{1},\mathbf{r}_{2}\right)$ and final
$\Psi_{f}^{-}\left(\mathbf{r}_{1},\mathbf{r}_{2}\right)$ states are
exact wave functions, or at least very good approximations for
them. Otherwise, differences between gauges are
apparent  \cite{ancarani_interplay_2008,kheifets_convergent_2004}.  For
simple systems such as He, these differences are restricted to the
magnitude of the cross sections, and minor deviations in the angular
distributions  \cite{Lucey98JPB} that are irrelevant for the
mainly qualitative study presented here. The calculation of the transition
matrix is performed by direct six-dimensional numerical integration in
the electronic spherical coordinates $\mathbf{r}_{1}$ and
$\mathbf{r}_{2},$ using a non-deterministic Vegas algorithm, with a
relative error is smaller than 3\% in the TDCS for all energies and angles
considered\cite{vegas}.

\begin{table}
  \caption{Energies and wave functions parameters (in atomic units) for ground and first excited
    states of He embedded in the model fullerene cage. }
  \centering \begin{tabular}{ccccccc}
    \toprule 
    $U_{0}$ & $E$ & N($\times$10$^{-2}$) & $\alpha$ & $a$ & $b$ & $c$\tabularnewline
    \hline
    \hline 
    \multicolumn{7}{c}{Ground State}\tabularnewline
    \hline
    1.10 & -2.8892 & 4.9390 &  0.47419 & -0.99835  & 0.00187  & 0.05723\tabularnewline
    1.35 & -2.9002  & 1.54985  & 0.54823 & -0.94505 & -0.31958 & 0.06879\tabularnewline
    1.60 & -3.0905  &  0.54473  &  0.61216  & -0.06843 & -0.99762 &  -0.00798\tabularnewline
    \hline 
    \multicolumn{7}{c}{First Excited State}\tabularnewline
    \hline
    1.10 & -2.7127 & 0.55145 & 0.47419  & 0.38490  & -0.91623 & 0.11117 \tabularnewline
    1.35 & -2.8879 & 4.71427 & 0.54823  & 0.99770 & -0.03930 & -0.05509\tabularnewline
    1.60 & -2.8890  & 4.94203 & 0.61216  & 0.99837 & -0.00609 & -0.05671\tabularnewline
    \bottomrule
  \end{tabular}
\end{table}

Let us turn our attention to the final state of the ionized electrons
in the continuum. Among many approximate wave functions for this
state, we choose a simple C3-style wave function (See
  \cite{otranto_kinetic_2003} and references
therein): \begin{equation}
  \Psi_{C3}^{cage}\left(\mathbf{r}_{1},\mathbf{r}_{2}\right)=\psi_{\mathbf{k}_{1}}^{-}(\mathbf{r}_{1})\psi_{\mathbf{k}_{2}}^{-}(\mathbf{r}_{2})
  D\left(\alpha_{12},\mathbf{k}_{12},\mathbf{r}_{12}\right)\end{equation}
where the electronic wave function in the combined field of the atomic
core ($-2/r$) and the fullerene cage ($V_{w}$, eq.(\ref{vcage})) is
described by the two-body functions
$\psi_{\mathbf{k}_{i}}^{-}(\mathbf{r}_{i})$ expanded in partial
waves. For comparison purposes, we also make use of a pure Coulomb
wave $\psi_{C3}^{Coul}(\mathbf{r}_{1},\mathbf{r_{2}})$, where the role
of the cage potential is neglected. The electron-electron correlation
is modeled with the usual Coulomb distortion factor
$D\left(\alpha_{12},\mathbf{k}_{12},\mathbf{r}_{12}\right)=
\;_{1}F_{1}\left(\alpha_{12},1,i\mathbf{k}_{12}\mathbf{r}_{12}+ik_{12}r_{12}\right)$
in terms of the Sommerfeld parameter $\alpha_{12}=1/k_{12}$ and the
relative momentum $\mathbf{k}_{12}=\mathbf{k}_{1}-\mathbf{k}_{2}$.
This model correctly reproduces the asymptotic condition of the
problem.  We recall that, although this model would not lead to
accurate absolute differential cross sections; it accounts for all
the features of the collisional process for this system.

We computed emission in an equal energy sharing situation
($E_{1}=E_{2}=10$ eV), with linearly polarized light. The calculations
of TDCS as a function of the magnitude of the fullerene cage $U_0$ are
displayed in Fig. 1. Overall, the cross sections interchange their
character as the states go through the avoided crossing, due to the
mirror collapse of the initial state\cite{mitnik_endohedrally_2008}. 
This is more evident for final Coulomb states (compare red lines in Fig. 1a
 with Fig. 1f; or Fig. 1d with Fig. 1c).

 The emission from the ground state below the crossing (Fig. 1d) fully
 agrees with the isolated atom DPI from a $1s^{2}$ state as expected,
 and is similar for both final stated used,
 $\psi_{C3}^{Coul}(\mathbf{r}_{1},\mathbf{r_{2}})$ or
 $\psi_{C3}^{cage}(\mathbf{r}_{1},\mathbf{r_{2}})$. Near and beyond
 the avoided crossing, the presence of the fullerene cage counteracts
 the interelectronic repulsion, moving the lobes towards the emission
 direction of electron 1 (Fig. 1d to Fig. 1f). This is slightly more
 important in the calculation that includes the cage in the final
 state (black curves in Fig. 1 online).

However, the excited state exhibits a more dramatic change along the crossing:
the cross section changes from the typical two-lobe configuration to a
four-lobe one. Thus, the mirror collapse observed in the intial state 
wave functions is broken in the cross sections. 
This effect can be clearly seen in Figs. 1a-c when compared with 1d-f, 
and it is present for both final states used, although it is noticeable at 
this energy only with the non-pure Coulomb final state 
$\psi_{C3}^{cage}(\mathbf{r}_{1},\mathbf{r_{2}})$ at this emission energy. 
Both the initial and final states contribute to this
feature. On one hand, the presence of a delocalized electron far from
the nucleus in the initial state slightly changes the interelectronic
repulsion. This has already been observed in calculations of double
photoionization from He(1s2s) states  \cite{colgan_total_2003}, and
were also attributed to the increasing extension of the initial state.
On the other hand, the introduction of the cage potential in the final
state $\psi_{C3}^{cage}(\mathbf{r}_{1},\mathbf{r_{2}})$ plays a very
important role, restraining the interelectronic repulsion (Fig. 1b) or
enhancing it (Fig. 1c) for different well depths. 

It is clear that the emission from the excited state deserves 
further investigation. The simple CI  that we adopted to describe the initial state
enables us to isolate the role of each CI state in the cross sections. We
computed the contribution of each CI state separately for the DPI of
the excited state, see Fig. 2. In this case, the energy sharing is
$E_{1}=E_{2}=50$ eV, and we choose a pure Coulomb wave
$\psi_{C3}^{Coul}(\mathbf{r}_{1},\mathbf{r_{2}})$ for the final
state. Contribution of the CI atomic states results in the typical
emission of electron 2 perpendicular to the direction of electron
1. However, the contribution of the atom-well CI state  presents a
complete different structure, with three clearly distiguishable
lobes. It is clear from this figure that the shape of the cross
section is mainly dictated by the \emph{interplay} between CI states, and
not by the simple superposition of them. This is clearly seen
comparing Fig. 2a and 2c, where the contribution of the coherent sum
determines the cross sections, while the role of the interference
among states clearly defines the behavior at the crossing, see Fig. 2b.

We can now compare the behavior of TDCS for different emission
energies.  While in Fig. 1b ($E_{1}=E_{2}=10$ eV) two lobes are well
defined when a pure Coulomb final wave is used, six lobes can be
observed in Fig2b for $E_{1}=E_{2}=50$ eV. This is not surprising,
since the electronic density of the atom-well CI state has a maximum
centered at the cage. Besides, partial waves with higher single
electron angular momentum contribute to the cross section for
increasing electron energies and are included
in the process, which results in these structures in the cross sections. 
This is a unique effect due to the presence of the cage,
because for isolated atoms, the centrifugal barrier inhibits the
penetration of high angular momenta partial waves into the region
where the atomic electronic density is significant.  We have checked
that this effect is enhanced when the final 
state $\Psi_{C3}^{cage}\left(\mathbf{r}_{1},\mathbf{r}_{2}\right)$ that
includes the fullerene cage is used.

In summary, we have computed double photoionization cross sections
from He endohedrically embedded in spherical fullerenes. This simple
system shows striking differences with the electronic emission from isolated atoms.
The presence of the fullerene cage determines the structure of the cross
sections, breaking the initial state mirror collapse, and enhancing
the role of higher angular momentum waves of the ejected electrons
in the process. This effect is more remarkable for higher energies,
although it can be seen for small ones, providing that the final state
also includes the cage potential. The relation between the present
results and other oscillatory properties found in this process deserves
further explorations.

F. D. C. would like to acknowledge the financial support of ANPCyT
(PICT 04/20548) and Conicet (PICT 5595). G. G. would like to
acknowledge the support by PICTR 03/0437 of the ANPCYT (Argentina) and
PGI 24/F038 of the Universidad Nacional del Sur (Argentina). D. M. acknowledges
the support by UBACyT X471, of the Universidad de Buenos Aires (Argentina).


\begin{thebibliography}{22}
  \expandafter\ifx\csname
  natexlab\endcsname\relax\def\natexlab#1{#1}\fi
  \expandafter\ifx\csname bibnamefont\endcsname\relax
  \def\bibnamefont#1{#1}\fi \expandafter\ifx\csname
  bibfnamefont\endcsname\relax \def\bibfnamefont#1{#1}\fi
  \expandafter\ifx\csname citenamefont\endcsname\relax
  \def  \citenamefont#1{#1}\fi \expandafter\ifx\csname
  url\endcsname\relax \def\url#1{\texttt{#1}}\fi
  \expandafter\ifx\csname urlprefix\endcsname\relax\def\urlprefix{URL
  }\fi \providecommand{\bibinfo}[2]{#2}
  \providecommand{\eprint}[2][]{\url{#2}}

\bibitem{kroto_c60:_1985} \bibinfo{author}{\bibfnamefont{H.~W.}
    \bibnamefont{Kroto}}, \bibinfo{author}{\bibfnamefont{J.~R.}
    \bibnamefont{Heath}}, \bibinfo{author}{\bibfnamefont{S.~C.}
    \bibnamefont{{O'Brien}}}, \bibinfo{author}{\bibfnamefont{R.~F.}
    \bibnamefont{Curl}}, \bibnamefont{and}
  \bibinfo{author}{\bibfnamefont{R.~E.} \bibnamefont{Smalley}},
  \bibinfo{journal}{Nature} \textbf{\bibinfo{volume}{318}},
  \bibinfo{pages}{162} (\bibinfo{year}{1985}).

\bibitem{bethune_atoms_1993} \bibinfo{author}{\bibfnamefont{D.~S.}
    \bibnamefont{Bethune}}, \bibinfo{author}{\bibfnamefont{R.~D.}
    \bibnamefont{Johnson}}, \bibinfo{author}{\bibfnamefont{J.~R.}
    \bibnamefont{Salem}}, \bibinfo{author}{\bibfnamefont{M.~S.}
    \bibnamefont{de~Vries}}, \bibnamefont{and}
  \bibinfo{author}{\bibfnamefont{C.~S.}  \bibnamefont{Yannoni}},
  \bibinfo{journal}{Nature} \textbf{\bibinfo{volume}{366}},
  \bibinfo{pages}{123} (\bibinfo{year}{1993}).

\bibitem{yannoni_scandium_1992} \bibinfo{author}{\bibfnamefont{C.~S.}
    \bibnamefont{Yannoni}},
  \bibinfo{author}{\bibfnamefont{M.}~\bibnamefont{Hoinkis}},
  \bibinfo{author}{\bibfnamefont{M.~S.} \bibnamefont{de~Vries}},
  \bibinfo{author}{\bibfnamefont{D.~S.} \bibnamefont{Bethune}},
  \bibinfo{author}{\bibfnamefont{J.~R.} \bibnamefont{Salem}},
  \bibinfo{author}{\bibfnamefont{M.~S.} \bibnamefont{Crowder}},
  \bibnamefont{and} \bibinfo{author}{\bibfnamefont{R.~D.}
    \bibnamefont{Johnson}}, \bibinfo{journal}{Science}
  \textbf{\bibinfo{volume}{256}}, \bibinfo{pages}{1191}
  (\bibinfo{year}{1992}).

\bibitem{pupysheva_fullerene_2008}
  \bibinfo{author}{\bibfnamefont{O.~V.} \bibnamefont{Pupysheva}},
  \bibinfo{author}{\bibfnamefont{A.~A.} \bibnamefont{Farajian}},
  \bibnamefont{and} \bibinfo{author}{\bibfnamefont{B.~I.}
    \bibnamefont{Yakobson}}, \bibinfo{journal}{Nano Letters}
  \textbf{\bibinfo{volume}{8}}, \bibinfo{pages}{767}
  (\bibinfo{year}{2008}).

\bibitem{footn1}{For a periodic table of endohedrally embedded atoms, see
  http://homepage.mac.com/jschrier/endo\-fullerenes\_table.html.}

\bibitem{puska_photoabsorption_1993}
  \bibinfo{author}{\bibfnamefont{M.~J.} \bibnamefont{Puska}}
  \bibnamefont{and} \bibinfo{author}{\bibfnamefont{R.~M.}
    \bibnamefont{Nieminen}}, \bibinfo{journal}{Phys. Rev. A}
  \textbf{\bibinfo{volume}{47}}, \bibinfo{pages}{1181}
  (\bibinfo{year}{1993}).

\bibitem{xu_oscillations_1996} \bibinfo{author}{\bibfnamefont{Y.~B.}
    \bibnamefont{Xu}}, \bibinfo{author}{\bibfnamefont{M.~Q.}
    \bibnamefont{Tan}}, \bibnamefont{and}
  \bibinfo{author}{\bibfnamefont{U.}~\bibnamefont{Becker}},
  \bibinfo{journal}{Phys. Rev. Lett.}
  \textbf{\bibinfo{volume}{76}}, \bibinfo{pages}{3538}
  (\bibinfo{year}{1996}).

\bibitem{amusia_strong_2000} \bibinfo{author}{\bibfnamefont{M.~Y.}
    \bibnamefont{Amusia}}, \bibinfo{author}{\bibfnamefont{A.~S.}
    \bibnamefont{Baltenkov}}, \bibnamefont{and}
  \bibinfo{author}{\bibfnamefont{U.}~\bibnamefont{Becker}},
  \bibinfo{journal}{Phys. Rev. A} \textbf{\bibinfo{volume}{62}},
  \bibinfo{pages}{012701} (\bibinfo{year}{2000}).

\bibitem{amusia_confinement_2004}
  \bibinfo{author}{\bibfnamefont{M.~Y.} \bibnamefont{Amusia}},
  \bibinfo{author}{\bibfnamefont{A.~S.} \bibnamefont{Baltenkov}},
  \bibinfo{author}{\bibfnamefont{V.~K.} \bibnamefont{Dolmatov}},
  \bibinfo{author}{\bibfnamefont{S.~T.} \bibnamefont{Manson}},
  \bibnamefont{and} \bibinfo{author}{\bibfnamefont{A.~Z.}
    \bibnamefont{Msezane}}, \bibinfo{journal}{Phys. Rev. A}
  \textbf{\bibinfo{volume}{70}}, \bibinfo{pages}{023201}
  (\bibinfo{year}{2004}).

\bibitem{dolmatov_correlation_2008}
  \bibinfo{author}{\bibfnamefont{V.~K.} \bibnamefont{Dolmatov}}
  \bibnamefont{and} \bibinfo{author}{\bibfnamefont{S.~T.}
    \bibnamefont{Manson}}, \bibinfo{journal}{J. Phys. B}
  \textbf{\bibinfo{volume}{41}}, \bibinfo{pages}{165001}
  (\bibinfo{year}{2008}).

\bibitem{amusia_photoionization_2008-1}
  \bibinfo{author}{\bibfnamefont{M.~Y.} \bibnamefont{Amusia}},
  \bibinfo{author}{\bibfnamefont{A.~S.} \bibnamefont{Baltenkov}},
  \bibnamefont{and} \bibinfo{author}{\bibfnamefont{L.~V.}
    \bibnamefont{Chernysheva}}, \bibinfo{journal}{J. Phys.
    B}
  \textbf{\bibinfo{volume}{41}}, \bibinfo{pages}{165201}
  (\bibinfo{year}{2008}).

\bibitem{kidun_multiple_2006}
  \bibinfo{author}{\bibfnamefont{O.}~\bibnamefont{Kidun}},
  \bibinfo{author}{\bibfnamefont{N.}~\bibnamefont{Fominykh}},
  \bibnamefont{and}
  \bibinfo{author}{\bibfnamefont{J.}~\bibnamefont{Berakdar}},
  \bibinfo{journal}{Comp. Mat. Science}
  \textbf{\bibinfo{volume}{35}}, \bibinfo{pages}{354}
  (\bibinfo{year}{2006}).

\bibitem{amusia_two-electron_2006}
  \bibinfo{author}{\bibfnamefont{M.~Y.} \bibnamefont{Amusia}},
  \bibinfo{author}{\bibfnamefont{E.~Z.} \bibnamefont{Liverts}},
  \bibnamefont{and} \bibinfo{author}{\bibfnamefont{V.~B.}
    \bibnamefont{Mandelzweig}}, \bibinfo{journal}{Phys. Rev. A}
  \textbf{\bibinfo{volume}{74}}, \bibinfo{pages}{042712}
  (\bibinfo{year}{2006}).

\bibitem{avaldi_photodouble_2005}
  \bibinfo{author}{\bibfnamefont{L.}~\bibnamefont{Avaldi}}
  \bibnamefont{and}
  \bibinfo{author}{\bibfnamefont{A.}~\bibnamefont{Huetz}},
  \bibinfo{journal}{J. Phys. B} \textbf{\bibinfo{volume}{38}},
  \bibinfo{pages}{S861} (\bibinfo{year}{2005}).

\bibitem{neek-amal_ground-state_2007}
  \bibinfo{author}{\bibfnamefont{M.}~\bibnamefont{{Neek-Amal}}},
  \bibinfo{author}{\bibfnamefont{G.}~\bibnamefont{Tayebirad}},
  \bibnamefont{and}
  \bibinfo{author}{\bibfnamefont{R.}~\bibnamefont{Asgari}},
  \bibinfo{journal}{J. Phys. B} \textbf{\bibinfo{volume}{40}},
  \bibinfo{pages}{1509} (\bibinfo{year}{2007}).

\bibitem{mitnik_endohedrally_2008}
  \bibinfo{author}{\bibfnamefont{D.~M.} \bibnamefont{Mitnik}},
  \bibinfo{author}{\bibfnamefont{J.}~\bibnamefont{Randazzo}},
  \bibnamefont{and}
  \bibinfo{author}{\bibfnamefont{G.}~\bibnamefont{Gasaneo}},
  \bibinfo{journal}{Phys. Rev. A} \textbf{\bibinfo{volume}{78}},
  \bibinfo{pages}{062501} (\bibinfo{year}{2008}).

\bibitem{connerade_electron_1999}
  \bibinfo{author}{\bibfnamefont{J.~P.} \bibnamefont{Connerade}},
  \bibinfo{author}{\bibfnamefont{V.~K.} \bibnamefont{Dolmatov}},
  \bibinfo{author}{\bibfnamefont{P.~A.} \bibnamefont{Lakshmi}},
  \bibnamefont{and} \bibinfo{author}{\bibfnamefont{S.~T.}
    \bibnamefont{Manson}}, \bibinfo{journal}{J. Phys. B}
  \textbf{\bibinfo{volume}{32}}, \bibinfo{pages}{L239}
  (\bibinfo{year}{1999}).

\bibitem{footn2}{This value for the crossing depends on the model used for
  the initial state.}

\bibitem{ancarani_interplay_2008}
  \bibinfo{author}{\bibfnamefont{L.~U.} \bibnamefont{Ancarani}},
  \bibinfo{author}{\bibfnamefont{G.}~\bibnamefont{Gasaneo}},
  \bibinfo{author}{\bibfnamefont{F.~D.} \bibnamefont{Colavecchia}},
  \bibnamefont{and} \bibinfo{author}{\bibfnamefont{C. Dal}
    \bibnamefont{Cappello}}, \bibinfo{journal}{Phys. Rev. A}
  \textbf{\bibinfo{volume}{77}}, \bibinfo{pages}{062712}
  (\bibinfo{year}{2008}).

\bibitem{kheifets_convergent_2004}
  \bibinfo{author}{\bibfnamefont{A.~S.} \bibnamefont{Kheifets}}
  \bibnamefont{and}
  \bibinfo{author}{\bibfnamefont{I.}~\bibnamefont{Bray}},
  \bibinfo{journal}{Phys. Rev. A} \textbf{\bibinfo{volume}{69}},
  \bibinfo{pages}{050701(R)} (\bibinfo{year}{2004}).

\bibitem{Lucey98JPB} \bibinfo{author}{\bibfnamefont{S.~P.}
    \bibnamefont{{Lucey}}},
  \bibinfo{author}{\bibfnamefont{J.}~\bibnamefont{{Rasch}}},
  \bibinfo{author}{\bibfnamefont{C.~T.} \bibnamefont{{Whelan}}},
  \bibnamefont{and} \bibinfo{author}{\bibfnamefont{H.~R.~J.}
    \bibnamefont{{Walters}}}, \bibinfo{journal}{J. Phys. B} 
\textbf{\bibinfo{volume}{31}},
  \bibinfo{pages}{1237} (\bibinfo{year}{1998}).

\bibitem{vegas} \bibinfo{author}{\bibfnamefont{T.}
    \bibnamefont{{Hahn}}},
     \bibinfo{journal}{Comp. Phys. Comm.} 
\textbf{\bibinfo{volume}{168}},
  \bibinfo{pages}{78} (\bibinfo{year}{2005}).


\bibitem{otranto_kinetic_2003}
  \bibinfo{author}{\bibfnamefont{S.}~\bibnamefont{Otranto}}
  \bibnamefont{and} \bibinfo{author}{\bibfnamefont{C.~R.}
    \bibnamefont{Garibotti}}, \bibinfo{journal}{Phys. Rev. A}
  \textbf{\bibinfo{volume}{67}}, \bibinfo{pages}{064701}
  (\bibinfo{year}{2003}).

\bibitem{colgan_total_2003}
  \bibinfo{author}{\bibfnamefont{J.}~\bibnamefont{Colgan}}
  \bibnamefont{and} \bibinfo{author}{\bibfnamefont{M.~S.}
    \bibnamefont{Pindzola}}, \bibinfo{journal}{Phys. Rev. A}
  \textbf{\bibinfo{volume}{67}}, \bibinfo{pages}{012711}
  (\bibinfo{year}{2003}).

\end{thebibliography}
%

\begin{figure*}[tp]
  \includegraphics[width=14cm]{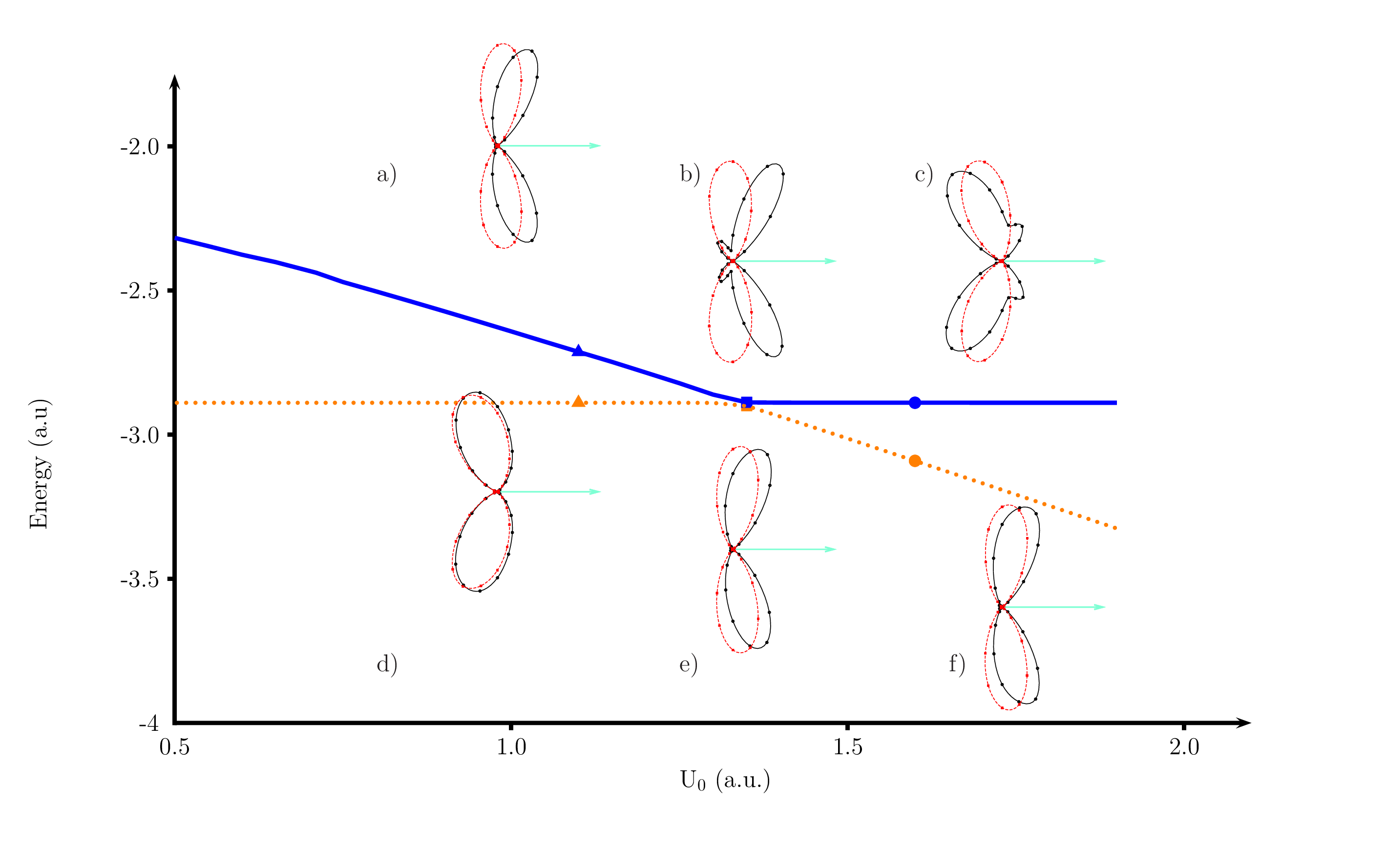}

  \caption{(Color online) Three fold differential cross section (TDCS)
    for $(\gamma,2e)$ ionization of the helium ground state in a
    fullerene cage, as a function of the angle of one of the ejected
    electrons $\theta_{2}$. The other one is ejected at  $\theta_{1}=0\degree$,
    fixed respect to the polarization $\mathbf{\varepsilon}$. The
    polarization vector is set along the $\mathbf{x}$ axis, and the
    impining light towards the page. The electrons are ejected with
    equal energy $E_{1}=E_{2}=10$ eV. Dotted (orange) and solid (blue)
    lines represent the ground and excited state energy as a function
    of $U_{0}$, respectively. TDCS are shown for both  excited (a, b, c) and
    ground states (d, e, f); and computed  with a pure Coulomb final state, dashed (red) line;
    or with exact potential including fullerene cage, solid (black) line. They
    are shown for $U_{0}=1.1$a.u. (triangles),  $U_{0}=1.35$a.u. (squares) and
    $U_{0}=1.6$ a.u. (circles). All TDCS are rescaled to one at maxima.}

  \label{Fig1}
\end{figure*}

\begin{figure}
  \includegraphics[width=17cm,clip]{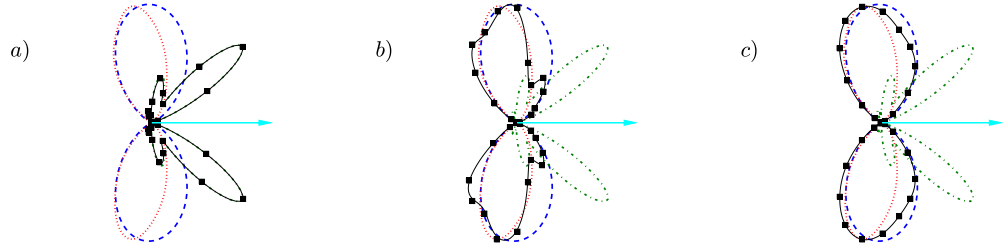}

  \caption{(Color Online) Contributions of the each initial CI states
    to the TDCS for $E_{1}=E_{2}=50$eV, calculated with the pure
    Coulomb final state for the DPI of the excited state of
    endohedrically embedded He. Kinematics as well as values of $U_0$
    are the same as in Fig. 1a)-1c). Solid (black online) line and
    squares: full TDCS, solid (cyan online) line, coherent
    contribution; dashed (blue) line, TDCS from first term, eq. (2);
    dash-dotted (green) line, TDCS from second term, eq (2); and
    dotted (red) line, TDCS from third term, eq. (2).All TDCS are rescaled to one at maxima.}

\end{figure}

\end{document}